\documentclass[12pt, letterpaper, epsfig]{article}

\usepackage{epsfig,lscape}
\usepackage{fancyhdr}	
\usepackage{amsmath, amsfonts}

\usepackage{natbib}
\usepackage{times}

\usepackage{enumerate}
\newtheorem{defn}{Definition}[section]

\newtheorem{thm}[defn]{Theorem}

\newtheorem{assu}[defn]{Assumption}

\newcommand{\cip}{\mbox{$\perp\!\!\!\perp$}}

\begin{document}

\title{Organic direct and indirect effects with post-treatment common causes of the mediator and the outcome}

\author{JUDITH J.~LOK$^\ast$\\
Department of Mathematics and Statistics\\
Boston University\\
MCS 232\\
111 Cummington Mall\\
Boston, MA 02215, USA\\
jjlok@bu.edu}

\markboth%
{Judith J. Lok}
{Organic direct and indirect effects}

\maketitle


\begin{abstract} Most of the literature on direct and indirect effects assumes that there are no post-treatment common causes of the mediator and the outcome. In contrast to natural direct and indirect effects, organic direct and indirect effects, which were introduced in Lok (2016, 2020), can be extended to provide an identification result for settings with post-treatment common causes of the mediator and the outcome. This article provides a definition and an identification result for organic direct and indirect effects in the presence of post-treatment common causes of mediator and outcome. These new organic indirect and direct effects have interpretations in terms of intervention effects. Organic indirect effects in the presence of post-treatment common causes are an addition to indirect effects through multivariate mediators. Organic indirect effects in the presence of post-treatment common causes can be used e.g.\ 1.\ to predict the effect of the initial treatment if its side affects are suppressed through additional interventions or 2.\ to predict the effect of a treatment that does not affect the post-treatment common cause and affects the mediator the same way as the initial treatment.
\end{abstract}

\noindent Causal inference, Direct and indirect effect,
  Mediation, Organic direct and indirect effect, Post-treatment common causes of mediator and outcome.

\section{Introduction}\label{introduction}

Most of the literature on direct and indirect effects assumes that there
are no post-treatment common causes of the mediator and the outcome (\cite{VanderWeelebook}). In contrast to natural direct and indirect effects, 
organic direct and indirect effects, introduced in \cite{medJL} and \cite{medJLBK}, can be extended to provide an identification result for settings with
post-treatment common causes of the mediator and the outcome, so-called mediator-outcome confounders. This
article provides a definition as well as an identification result for organic direct and
indirect effects in the presence of post-treatment common causes of
the mediator and the outcome. This provides another
alternative to the three quantities described in \cite{vanderweele2014effect}. 

Organic direct and indirect effects from \cite{medJL} and \cite{medJLBK} are intervention based approaches. Indirect effects are the effects of organic interventions.
It is important to differentiate between interventions on $M$ that
happen before $L$ and interventions on $M$ that happen after $L$, because the distribution of $L$ under
an intervention depends on when that intervention takes place. 
If one is after interventions that also affect $L$, it will often be more informative to consider $L$ as part of a multivariate mediator $(M,L)$. Causal mediation analysis proposed in e.g.\ \cite{RobGreenmed,Pearlmed,VanderWeelebook,Imai,Ericmedsurv,Tyler,medJL,medJLBK} includes such multivariate mediators  $(M,L)$. Hence, this article only considers organic interventions $I$ on the mediator
that happen after $L$. This is of particular importance when we evaluate e.g.\ the effect of interventions that reduce a side effect of the treatment $A=1$. 

Just as organic direct and indirect effects in the absence of post-treatment common causes of the mediator and the outcome, the proposed organic direct and indirect effects do not require that the mediator can be set to any specific value. It suffices that there are ``organic'' interventions
on the mediator that change its distribution.

We provide two alternative definitions, following \cite{medJL} and \cite{medJLBK}. The first definition is closer to natural direct and indirect effects (\cite{RobGreenmed,Pearlmed,VanderWeelebook,Imai,Ericmedsurv,Tyler}), and extends \cite{medJL}. The second definition is closer to \cite{BaronKenny}, and extends \cite{medJLBK}.

\section{Setting and notation}

Denote treatment by $A$, the
mediator by $M$, the outcome by $Y$, pre-treatment common causes of the mediator and the outcome by
$C$, and post-treatment common causes of the mediator and the outcome by $L$. Following \cite{medJL,medJLBK}, $I$ indicates an intervention on the
mediator $M$. Initially, I assume that treatment $A$ is randomized; this assumption can be relaxed as in \cite{medJL,medJLBK}. For the current setting,
the DAG is shown in Figure~1.\\

\begin{figure}[htb!]
\begin{picture}(400,165)
\thicklines
\put(230,165){\makebox(0,0){Figure 1: DAG summarizing the data in the presence of a post-treatment}}
\put(198,150){\makebox(0,0){common cause of mediator and outcome $L$}}
\put(70,55){\makebox(0,0){$A$}}
\put(175,55){\makebox(0,0){$L$}}
\put(285,55){\makebox(0,0){$M$}}
\put(390,55){\makebox(0,0){$Y$}}
\put(285,5){\makebox(0,0){$C$}}
\thicklines
\put(90,53){\vector(1,0){65}}
\put(195,53){\vector(1,0){70}}
\put(305,53){\vector(1,0){65}}
\put(285,12){\vector(-3,1){95}}
\put(285,12){\vector(0,1){33}}
\put(285,12){\vector(3,1){95}}
\qbezier(80,60)(230,215)(375,60)
\qbezier(80,60)(175,110)(274,60)
\qbezier(185,60)(285,160)(370,60)
\put(370,60){\vector(2,-1){3}}
\put(375,60){\vector(2,-1){3}}
\put(275,60){\vector(2,-1){3}}
\end{picture}
\end{figure}

\noindent
From the DAG notice that an intervention on the mediator $M$ may happen
after $L$, because $L$ is realized before $M$.

Organic direct and indirect effects from \cite{medJL} and \cite{medJLBK} are an intervention based approaches. 
It is important to differentiate between interventions on $M$ that
happen before $L$ and after $L$, because the distribution of $L$ under
the intervention depends on when the intervention takes place. 
This article only considers organic interventions $I$ on the mediator
that happen after $L$. 

A subscript $1$ indicates ``under treatment, $A=1$'', and a subscript $0$ indicates ``under no treatment, $A=0$''. $M_{1,I=1}$ and $Y_{1,I=1}$ indicate the mediator and the outcome under treatment, $A=1$, and organic intervention $I$. $M_{0,I=1}$ and $Y_{0,I=1}$ indicate the mediator and the outcome under no treatment, $A=0$, and organic intervention $I$.

\section{Definition, identifiability and estimation of organic direct and indirect effects with post-treatment common causes of mediator and outcome: interventions after $A=1$}\label{posttrtA1}

If $I$ happens after the post-treatment common cause $L$ of the mediator and the outcome, the value of $L$ under the intervention $I$ and
treatment $A=1$ equals the value of $L_1$. Hence, we define
\begin{defn}\emph{(Organic intervention in the presence of post-treatment common causes $L$ of mediator and outcome).}\label{mimpost}
An intervention $I$ is an organic intervention if for all $l$, $c$,
\begin{equation}\label{defintpost}
M_{1,I=1}|L_1=l,C=c \sim M_0 |L_0=l,C=c
\end{equation}
and
\begin{equation}\label{identpost}
Y_{1,I=1}|M_{1,I=1}=m,L_1=l,C=c \sim Y_1|M_1=m,L_1=l,C=c.
\end{equation}
\end{defn}
The idea behind mediation analysis is that under the intervention $I$ and
treatment, $M_{1,I=1}$ resembles $M_0$. Therefore, this article assumes
that $M_{1,I=1}$ depends on $L$ and $C$ in the same way as $M_0$, just as
without $L$, \cite{medJL} assumed that $M_{1,I=1}$ depends on $C$ in the
same way as $M_0$. Notice that without $L$, Definition~\ref{mimpost}
simplifies to Definition~4.1 from \cite{medJL}. Notice also that in contrast to \cite{medJL},
$M_{1,I=1}=M_0$, the basis for natural direct and indirect effects, is no longer a special case of an organic intervention.

Equation~(\ref{identpost}) means that given $L_1=l$ and $C=c$, the
prognosis under treatment of a unit ``with $M_{1,I=1}=m$'' is the same as
the prognosis under treatment of a unit ``with $M_1=m$''. In other
words, given $C$ and $L_1$, treated units with observed mediator equal
to $m$ are representative of treated units with $M_{1,I=1}=m$. Similar
to \cite{medJL}, equation~(\ref{identpost}) can be relaxed to
\begin{equation*}E\left[Y_{1,I=1}|M_{1,I=1}=m,L_1=l,C=c\right]
=E\left[Y_1|M_1=m,L_1=l,C=c\right].
\end{equation*}

The consistency assumption is straightforward, but needs to include $L$:
\begin{assu} \emph{(Consistency).} \label{conspost} If $A=1$, then $L=L_1$, $M=M_1$ and $Y=Y_1$. If $A=0$, then $L=L_0$, $M=M_0$ and $Y=Y_0$.
\end{assu}

The following identification result holds:
\begin{thm}\label{thmpost}\emph{(Organic direct and indirect effects: identification in the presence of post-treatment common causes $L$ of mediator and outcome).} Under randomized treatment, consistency assumption~\ref{conspost} and definition of organic interventions~\ref{mimpost}, $E\left(Y_{1,I=1}\right)$, for an organic intervention $I$, is equal to
\begin{equation*}
\int_{(c,l,m)}E\left[Y|M=m,L=l,C=c,A=1\right]f_{M|L=l,C=c,A=0}(m)f_{L|C=c,A=1}(l)f_C(c)dm\,dl\,dc.
\end{equation*}
\end{thm}
All objects on the right hand side of the equation in Theorem~\ref{thmpost} depend on observables only and can be fitted using standard methods. Inference follows along the lines of Section~6 of \cite{medJL}, see also Section~\ref{inferencepost} below.

\noindent {\bf Proof of theorem~\ref{thmpost}}
\begin{eqnarray*}
\lefteqn{E\left(Y_{1,I=1}\right)
=E\left(E\left[Y_{1,I=1}|M_{1,I=1},L_1,C\right]\right)}\\
&=&\int_{(c,l,m)}E\left[Y_{1,I=1}|M_{1,I=1}=m,L_1=l,C=c\right]f_{M_{1,I=1}|L_1=l,C=c}(m)dm\,f_{L_1|C=c}(l)dl\,f_C(c)dc\\
&=&\int_{(c,l,m)}E\left[Y_{1}|M_1=m,L_1=l,C=c\right]f_{M_0|L_0=l,C=c}(m)dm\,f_{L_1|C=c}(l)dl\,f_C(c)dc\\
&=&\int_{(c,l,m)}E\left[Y_{1}|M_1=m,L_1=l,C=c,A=1\right]f_{M_0|L_0=l,C=c,A=0}(m)dmf_{L_1|C=c,A=1}(l)dl f_C(c)dc\\
&=&\int_{(c,l,m)}E\left[Y|M=m,L=l,C=c,A=1\right]f_{M|L=l,C=c,A=0}(m)dm\,f_{L|C=c,A=1}(l)dl\,f_C(c)dc.
\end{eqnarray*}
The first two equalities follow from the definition of
conditional expectation. The third equality follows from
equations~(\ref{identpost}) and~(\ref{defintpost}). The fourth
equality follows from the fact that treatment was randomized; this
implies that
\begin{equation*}A\cip \left(Y_{1},M_1,L_1\right)|C \hspace{1cm}{\rm and}\hspace{1cm}A\cip \left(M_0,L_0\right)|C.
\end{equation*}
The last equality follows from Consistency Assumption~\ref{conspost}.
\hfill $\Box$

\section{Definition, identifiability and estimation of organic direct and indirect effects with post-treatment common causes of mediator and outcome: interventions after $A=0$}\label{posttrtA0}

In this section we start with $A=0$, and evaluate the effect of an intervention $I$ that happens after $L$ but creates a setting where the distribution of the mediator given $L$ resembles that of the mediator given $L$ under treatment $A=1$. This setting is relevant e.g.\ if interest lies in a treatment that does not affect $L$ but has a similar effect on the mediator as $A=1$. In this case, the value of $L$ under the intervention and $A=0$ equals the value of $L_0$. We define organic interventions in
this setting as follows:
\begin{defn}\emph{(Organic intervention in the presence of post-treatment common causes $L$ of mediator and outcome).}\label{mimpost0}
An intervention $I$ is an organic intervention if for all $l$, $c$,
\begin{equation}\label{defintpost0}
M_{0,I=1}|L_0=l,C=c \sim M_1 |L_1=l,C=c
\end{equation}
and
\begin{equation}\label{identpost0}
Y_{0,I=1}|M_{0,I=1}=m,L_0=l,C=c \sim Y_0|M_0=m,L_0=l,C=c.
\end{equation}
\end{defn}
The idea behind mediation analysis is that under the intervention $I$ and no
treatment, $M_{0,I=1}$ resembles $M_1$. Therefore, this article assumes
that $M_{0,I=1}$ depends on $L$ and $C$ in the same way as $M_1$, just as
without $L$, \cite{medJLBK} assumed that $M_{0,I=1}$ depends on $C$ in the
same way as $M_1$. Without $L$, Definition~\ref{mimpost0}
simplifies to \cite{medJLBK}. As in \cite{medJLBK}, $M_{0,I=1}=M_1$ is not a special case of an organic intervention.

Equation~(\ref{identpost0}) means that given $L_0=l$ and $C=c$, the
prognosis under ``no treatment'' of a unit ``with $M_{0,I=1}=m$'' is the same as
the prognosis under ``no treatment'' of a unit ``with $M_0=m$''. In other
words, given $C$ and $L_0$, untreated units with observed mediator equal
to $m$ are representative of untreated units with $M_{0,I=1}=m$. Similar
to \cite{medJL,medJLBK}, equation~(\ref{identpost0}) can be relaxed to
\begin{equation*}
E\left[Y_{0,I=1}|M_{0,I=1}=m,L_0=l,C=c\right]
=E\left[Y_0|M_0=m,L_0=l,C=c\right].
\end{equation*}

The consistency assumption is the same as Consistency Assumption~\ref{conspost}. The following identification result follows from the same proof as Theorem~\ref{thmpost}:
\begin{thm}\label{thmpost0}\emph{(Organic direct and indirect effects: identification in the presence of post-treatment common causes $L$ of mediator and outcome).} Under randomized treatment, consistency assumption~\ref{conspost} and definition of organic interventions~\ref{mimpost}, $E\left(Y_{0,I=1}\right)$, for an organic intervention $I$, is equal to
\begin{equation*}
\int_{(c,l,m)}E\left[Y|M=m,L=l,C=c,A=0\right]f_{M|L=l,C=c,A=1}(m)f_{L|C=c,A=0}(l)f_C(c)dm\,dl\,dc.
\end{equation*}
\end{thm}

\section{Organic direct and indirect effects with post-treatment common causes of mediator and outcome: interventions after $A=1$}\label{inferencepost}

In the presence of post-treatment common causes $L$ of the mediator
and the outcome, inference can be done based on Section~\ref{posttrtA1},
Theorem~\ref{thmpost}.  For example, suppose that
\begin{equation}\label{m1m0post}
M_1\sim M_0+\beta_1+\beta_4C+\beta_5L|C,L
\end{equation}
with $\beta_1\in\mathbb{R}$, $\beta_4\in\mathbb{R}^k$, and 
would be the case if, for example,
\begin{equation*}
M=\beta_0+\beta_1A+\beta_2C+\beta_3L+\beta_4AC+\beta_5AL+\beta_6CL+\epsilon,
\end{equation*}
where the random variable $\epsilon$ has the same distribution given
$(C,L)$ under treatment as without treatment, and with
$\beta_0,\beta_1\in\mathbb{R}$,
$\beta_2,\beta_4\in\mathbb{R}^k$, $\beta_3,\beta_5\in\mathbb{R}^l$, and $\beta_6\in\mathbb{R}^{p}$. Suppose in
addition that the expected value of $Y$ given $M$, $L$, and $C$ under
treatment follows some parametric model of the form
\begin{equation}\label{Yparpost}
E\left[Y|M=m,L=l,C=c,A=1\right]=f_\theta(m,l,c).
\end{equation}
Then, Theorem~\ref{thmpost}
implies that
\begin{eqnarray}\label{expexprpost}\lefteqn{E\left(Y_{1,M_{1,I=1}}^{I}\right)}\nonumber\\
&=&\int_{(m,l,c)}E\left[Y|M=m,L=l,C=c,A=1\right]f_{M|L=l,C=c,A=0}(m)f_{L|C=c,A=1}(l)f_C(c)dmdldc\nonumber\\
&=&\int_{(c,l,m)}f_\theta(m,l,c)f_{M|L=l,C=c,A=1}(m+\beta_1+\beta_4c+\beta_5l)f_{L|C=c,A=1}(l)f_C(c)dmdldc\nonumber\\
&=&\int_{(c,l,\tilde{m})}f_\theta(\tilde{m}-\beta_1-\beta_4c-\beta_5l,l,c)f_{M|L=l,C=c,A=1}(\tilde{m})f_{L|C=c,A=1}(l)f_C(c)d\tilde{m}dldc\nonumber\\
&=&E\left[f_\theta(M-\beta_1-\beta_4C-\beta_5L,L,C)|A=1\right],
\end{eqnarray}
just as in Section~6 of \cite{medJL}.
Expression (\ref{expexprpost}) can
be estimated by fitting models (\ref{m1m0post}) and (\ref{Yparpost})
above using standard methods, plugging the parameter estimates in
(\ref{expexprpost}), and replacing the expectation given $A=1$ by its
empirical average. Standard errors can be estimated using the bootstrap.

\section{Discussion}

This article provides an intervention based approach to mediation analysis, extended to settings with post-treatment common causes $L$ of the mediator $M$ and the outcome $Y$. The types of interventions that are relevant depend on the subject matter, on what one intends to do with the results of the mediation analysis.

If interest lies in interventions that affect not only the mediator $M$ but also the post-treatment common causes $L$, mediation analysis with a multivariate mediator $(M,L)$ could be most relevant. Existing mediation analysis approaches from e.g.\  \cite{RobGreenmed,Pearlmed,VanderWeelebook,Imai,Ericmedsurv,Tyler,medJL,medJLBK} can be used for the multivariate mediator $(M,L)$.

If interest lies in interventions that affect only the mediator $M$ (and not $L$), the approach in this article, which extends \cite{medJL} and \cite{medJLBK} to settings with post-treatment common causes $L$ of the mediator $M$ and the outcome $Y$, can be used. In our approach, interventions are considered that do not affect the post-treatment common cause $L$. Organic interventions $I$ affect the distribution of the mediator $M$ given the common causes $(L,C)$ so that $M$ follows the distribution of the mediator given the ``other'' treatment option. That is, when combining an organic intervention $I$ with $A=1$, the distribution of the mediator $M_{1,I=1}$ given $(L=l,C=c)$, which now equals $(L_1=l,C=c)$, is affected to become the distribution of the mediator $M_0$ given $(L_0=l,C=c)$. And, when combining an organic intervention $I$ with $A=0$, the distribution of the mediator $M_{0,I=1}$ given $(L=l,C=c)$, which now equals $(L_0=l,C=c)$, is affected to become the distribution of the mediator $M_1$ given $(L_1=l,C=c)$. After that, the outcomes follow their ``natural course'', that is, $Y_{1,I=1}$ continues to evolve as $Y_1$ given $(M_1,L_1,C)$ and $Y_{0,I=1}$ continues to evolve as $Y_0$ given $(M_0,L_0,C)$.

Settings where interest lies in interventions that affect only the mediator $M$ (and not $L$) include settings where interventions on the mediator are designed to counter side-effects of a treatment $A=1$. They also include settings with new treatments which are designed to not affect the post-treatment common cause $L$ but which do affect the mediator $M$.

Our proposed organic direct and indirect effects add another decomposition of the treatment effect into effects through different pathways, adding to existing methods described in \cite{vanderweele2014effect}. We show that organic direct and indirect effects are identifiable in settings with post-treatment common causes of the mediator and the outcome. We also provide examples of inference. Organic direct and indirect effects have interpretations in terms of intervention effects, and thus contribute to critical thinking about the purpose of mediation analyses.

\addcontentsline{toc}{chapter}{Bibliography}
\bibliographystyle{chicago} \bibliography{ref}

\end{document}